# A Comparison of the Galactic Cosmic Ray H, He and C/O Nuclei Spectra Measured Between ~5 and 500 MeV/nuc Beyond 122 AU at Voyager 1 with the Predictions of a Diffusion Model for Propagation in the Galaxy

W.R. Webber[1], P.R. Higbie[2] and F.B. McDonald[3+]

1. New Mexico State University, Department of Astronomy, Las Cruces, NM 88003, USA

2. New Mexico State University, Physics Department, Las Cruces, NM 88003, USA

3. University of Maryland, Institute of Physical Science and Technology, College Park, MD 20742, USA

+ Deceased August 31, 2012

**ABSTRACT**

After the disappearance of lower energy heliospheric particles at Voyager 1 starting on August 25$^{th}$, 2012, spectra of H, He and C/O nuclei were revealed that resembled those to be expected for galactic cosmic rays. These spectra had intensity peaks in the range of 30-60 MeV, decreasing at both lower energies down to a few MeV and at higher energies up to several hundred MeV. We have modeled the propagation of these particles in the galaxy using an updated Leaky Box Diffusion model which determines the spectra of these components from ~2 MeV to >200 GeV. The key parameters used in the model are a galactic input spectrum ~$P^{-2.24}$, the same for all components and independent of rigidity, and a diffusion coefficient that is ~$P^{0.5}$ above a lower rigidity and increases ~$\beta^{-1.0}$ below a lower rigidity ~0.56 GV. These same parameters also fit the high energy H and He data from ~10-200 GeV/nuc from the PAMELA and BESS experiments. The new Voyager spectra for all three nuclei are thus consistent with rigidity spectra ~$P^{-2.24}$ from the lowest energies to at least 100 GeV. Deviations from this spectrum can reasonably be attributed to propagation effects. Some deviations between the calculated and newly observed spectra are noted, however, below ~30 MeV/nuc, particularly for C/O nuclei, that could be significant regarding the propagation and sources of these particles.

# 1. Introduction

Until recently the spectra of low energy galactic cosmic ray nuclei (GCR) below ~100 MeV were obscured by solar modulation effects in the heliosphere and the presence of anomalous cosmic rays (ACR) presumably accelerated in the heliosheath. However, starting on August 25$^{th}$, 2012, the ACR rapidly disappeared at Voyager 1, at a distance of 121.7 AU. Within a few weeks the intensities of ACR decreased by a factor of over 100 revealing spectra of what appear to be GCR H, He and C/O primary nuclei (Webber and McDonald, 2013; Stone, et al., 2013). These GCR spectra were more intense than those observed earlier in 2012 indicating that significant "solar" modulation of GCR occurred at the same time the ACR decreased. This modulation between ~120-122 AU is described in more detail in Webber, Higbie and McDonald, 2013.

The H, He and C/O nuclei spectra that are observed near the end of 2012 at V1 have differential spectra that are decreasing in intensity below ~30-50 MeV (where the spectral intensity peak occurred). These GCR spectra look much like previous estimates of galactic spectra that had been determined in galactic diffusion models where the effects of an increasing diffusion coefficient at low rigidities as well as ionization energy loss in the interstellar medium can cause a spectral turnover at low energies (Ip and Axford, 1968; Putze, Maurin and Donato, 2011; Webber and Higbie, 2009; Strong and Moskalenko, 1998).

In the previous 6 years before 2012 the spectra of these cosmic ray nuclei unfolded dramatically with the lower energies increasing rapidly as V1 moved through the outer heliosheath from ~100-120 AU (Webber and Higbie, 2009; Webber, 2012). In fact, by the beginning of 2012 these spectra had unfolded (increased) to the point where any "residual" solar modulation potential beyond ~120 AU necessary to produce the spectra observed at the end of 2011 would have to be ~80 MV or less in a Parker like (1958) diffusion-convection modulation model, where the total modulation potential between the Earth and interstellar space was ~270 MV at that time (Webber, Higbie and McDonald, 2013).

Just before the ACR disappeared at 2012.65, the lowest energy GCR that could be measured without a severe ACR background correction were ~60 MeV/nuc He nuclei. These He particles increased by a factor ~3 over a time period from early in 2012 to after 2012.65, consistent with a large and sudden decrease in the remaining modulation potential. It is the significance of these low energy and low level of modulation spectra of GCR that were observed at V1 at the end of 2012, after the final intensity increase, and how these spectra relate to the properties of GCR propagation in the galaxy at low and high energies that we wish to discuss in this paper.

## 2. The Spectra Measured at V1 at the end of 2012

In Figure 1 we show the H nuclei spectrum between 2 and 400 MeV measured by V1 between 09/16/2012 and the end of 2012. Figure 2 show the He nuclei spectra between 2 and 600 MeV/nuc for the same time period. There was no discernible change of intensities during this 3.5 month time period. In Figure 3 we show the C/O nuclei spectrum between ~5 and 120 MeV/nuc. All of these spectra are from Stone, et al., 2013. There is no evidence of any substantial heliospheric anomalous cosmic ray background in the H and He spectra, down to ~3 MeV/nuc.

## 3. Calculation of Propagated Spectra in the Galaxy

This paper represents an extension of the galactic propagation calculations of Webber and Higbie, 2009. These earlier calculations, which were instigated to help understand measurements of H and He spectra at V1 out to ~105 AU, used two types of diffusion models to calculate the propagated interstellar spectra of cosmic ray H, He and C/O nuclei at the boundary of the heliosphere. The first model was a Monte Carlo diffusion model (MCDM). The second model was a variant of the simple but widely used Leaky box model (LBM) which was earlier used to interpret Voyager spectral data on He and heavier nuclei in 1998 when V2 was at 56 AU (Webber, McDonald and Lukasiak, 2003). These LBM calculations are very similar in concept to the comprehensive calculations using a diffusion model for nuclei made by Putze, Maurin and Donato, 2012.

Our comparison of the two propagation models at that time showed that, for essentially equivalent propagation parameters and source spectra, the MCDM and LBM models gave nearly identical propagated spectra above ~100 MeV/nuc. At lower energies the LBM is to be preferred because of several limitations of the MCDM. Also the appropriate propagation parameters and cross sections are available for use in this model to extend the calculations down to 10 MeV/nuc and below. For this reason we use here the LBM for the propagation calculations.

The LBM has been discussed in several publications (Webber, McDonald and Lukasiak, 2003; Webber and Higbie, 2009). It is a modified LBM originally developed by the French group (Engelmann, et al., 1990). The interstellar medium is assumed to be 90% H and 10% He and to have an average density =0.4 atoms/cm$^3$. The full complement of cross sections for H and He interstellar matter from Webber, et al., 2003, are used. Added ionization energy loss due to 15% ionized hydrogen is included. The parameters for the propagation model are adjusted to fit both high and low energy data on the C/O intensities and also the B/C and N/O ratios using source rigidity spectra $\sim P^{-2.2}$-$P^{-2.3}$ with the result that a path length $\lambda = \sim 15.1 \, \beta(P/P_0)^{0.5}$ g/cm$^2$ above a rigidity $P_0$ has been obtained.

In Webber and Higbie, 2009, the local interstellar spectra (LIS) above ~100 MeV were calculated for several possibilities for the variation of the interstellar diffusion coefficient below $P_0$= 3 GV and for source spectra which were assumed to be of the form dj/dp $\sim P^{-x}$ where P is rigidity and x is the source spectrum has a value between -2.1 and -2.4. This gave propagated spectra that had exponents = -(x + 0.5) =-(2.6-2.9) at high rigidities (>10 GV). It was found from these studies that the value of the diffusion coefficient and its rigidity dependence was a dominant determinator of the calculated spectral shape of the LIS at low energies, more important above ~30 MeV/nuc than the value of the exponent of the source spectrum and other effects such as ionization energy loss. The new Voyager data therefore provides a very sensitive test of possible changes in the galactic diffusion coefficient at rigidities ~3 GV and below of the type that were originally suggested by Ptuskin, et al., 2006 (see also Strong and Moskalenko, 1998).

The diffusion coefficient rigidity dependences used in these earlier calculations are shown in Figure 4. Curve (1) was used in the LBM calculation and Curve (2) in the MCDM calculation. The resulting LIS for H, He and C/O nuclei from these earlier calculations are shown in Figures 1, 2 and 3 along with the recent spectra of these nuclei measured by Voyager 1 beyond 121.7 AU (Stone, et al., 2013). It is seen that the calculated spectra for the source exponent x=-2.3 for the MCDM and LBM models actually bracket the observed spectra measured by V1 at the end of 2012 by Stone, et al., 2013, above ~100 MeV/nuc and are closer to those of the LBM and its choice of diffusion coefficient which was taken to be a constant below ~3 GV. The earlier calculations, however, did not extend to the much lower energies now being observed by V1. The object of the present paper is therefore to calculate the spectra observed by Voyager 1 in late 2012 within the concepts and parameters of the LBM being used.

The new calculations use all of the original propagation parameters and cross sections used in Webber and Higbie, 2009, and include updates in the cross section values to extend them below 100 MeV/nuc. These include the Letaw, Silberberg and Tsao, 1983, total cross section formula below ~100 MeV/nuc. The updates also include measurements (of which there are few) and the individual charge changing cross sections from the Fluka-CERN calculations. These were the individual cross sections used in the calculation of 7Be and 10Be production in the Earth's atmosphere (Webber and Higbie, 2009). Because of the large secondary production of 3He and 2H, amounting to >20% of 4He at 1 GeV/nuc, the 3He and 4He and 1H and 2H propagation was calculated separately and the results added to obtain the total He and H abundances.

In these new calculations the source exponent of the rigidity spectra of the nuclei was varied from ~$P^{-2.2}$ to ~$P^{-2.4}$ and the galactic diffusion coefficient was taken to be ~$P^{0.5}$ above a certain lower rigidity $P_0$ which is taken to vary between 0.32-3.0 GV. These parameters therefore would result in spectra at high energies (>10 GeV/nuc) which are ~$P^{-2.70}$ to $P^{-2.90}$.

Below $P_0$ the diffusion coefficient was taken to have dependence of between $\beta^{-1.0}$ and $\beta^{-2.0}$. Examples of this diffusion coefficient for a $\beta^{-1.0}$ dependence and for $P_0$=0.32, 0.56 and 1.00 GV are shown in Figure 4 for the new calculations.

The source spectral exponent was assumed to be the same at all rigidities and for all three charges. This model assumes any convection effects or re-acceleration equal to zero.

After a large number of calculations, propagated spectra that give a good fit to the H, He and C/O nuclei spectra above a few MeV and up to the highest energies are shown in Figures 5, 6 and 7 respectively for H, He and C/O nuclei. These calculations use "source" spectra ~$P^{-2.24}$ and values of $P_0$ between 0.32 and 1.00 GV and a ~$\beta^{-1.0}$ dependence of the diffusion coefficient at low rigidities.

All figures are "split" figures in which the data below 1 GeV/nuc is shown in a normal log vertical scale of intensity whereas above 1 GeV/nuc the differential intensity is multiplied by $E^{2.5}$. This is done to compare the intensity differences between calculations and observations at both low energies (~10-100 MeV/nuc) and high energies (~10-100 GeV/nuc) with essentially the same expanded vertical scale.

Considering the lower energies first, we note that the new V1 data are well fit for all three nuclei, H, He and C/O with a value of $P_0=0.56 \pm 0.24$ GV and the source spectral exponent = -2.24. Below ~30 MeV/nuc there are differences between predictions and measurements for all of the spectra. At these lower energies the measured intensities lie below the predictions. This is most evident for the C/O nuclei spectrum.

The low energy data of Stone, et al., 2013, could also be fit using values of $P_0=0.32$ and 1.0 GV but this would require changing either the spectral exponent or the absolute intensity normalization, or both. The lowest least squares fitting variations were obtained for $P_0$ between 0.56 and 0.76 GV.

The high energy observations of the intensity for these nuclei play a very important role. Note that, for the $P^{-2.24}$ spectra and the normalization required to fit the low energy data, the intensities at high energies ~10 GeV/nuc and above, where the solar modulation becomes negligible for all three components, are also fit to within $\pm 10\%$. This would not be the case if the spectra were changed to $P^{-2.20}$ or $P^{-2.30}$ and the normalizations were changed to fit the low energy data. In fact this ability to fit to within ~10% the data at <u>both</u> low and high energies using $P^{-2.24}$ spectra with the exponent independent of rigidity over a factor ~100-200 in rigidity implies that the source spectra of all three nuclei are very similar and this "source" spectral exponent is unchanging as a function of P and equal to $P^{-2.24}$ to an accuracy of $\pm 0.05$ in the exponent in this rigidity range. The changes in the spectra that are observed are therefore almost completely explained (above ~30 MeV/nuc) using the parameters of the propagation model. This is a key conclusion of this paper.

The new PAMELA data on H and He nuclei (Adriano, et al., 2013, Potgieter, et al., 2013), which are shown along with BESS data in Figures 5 and 6 are very relevant to this discussion and the normalization. These PAMELA H and He spectra at the Earth in 2009 require an overall modulation potential at the Earth of ~270 MV using the H and He spectra measured by V1 as the low energy part of the LIS spectra along with a "Parker" type diffusion-convection modulation model to describe heliospheric modulation appropriate to the PAMELA data (see also Webber, Higbie and McDonald, 2013).

## 4. General Discussion and Implication of the Voyager H, He and C/O Spectra

It is perhaps remarkable that the detailed and statistically accurate, to within less than ±10%, Voyager H, He and C/O nuclei spectra observed at low energies in late 2012 can be predicted to an accuracy of a few percent between ~4 and 400 MeV/nuc and then extended to higher energies as well using a simple galactic propagation model which assumes all of the relevant parameters and processes(diffusion, ionization energy loss and a simple source spectral exponent), as well as the source distribution itself, are more or less uniform in what is obviously a very complex galaxy in its spatial features. This cosmic ray "fog", which is due to diffusion and the uniformness of the source distribution, is essentially so complete that not a single observational feature is preserved that could define the spatial features or source distribution of any of the parameters involved in the acceleration and propagation process. The seminal article by Schlickeiser and Lerche, 1985, entitled, "Why does the Leaky Box model work so well", helps to explain why. In the end the answer lies in the fact that the age distributions as observed in the LIS, are exponentials.

But there are some deviations between the Voyager observations and the galactic propagation models as shown in Figures 5, 6 and 7 that may need further study. Below about ~10 MeV for H nuclei the Voyager observations lie below the predictions. For He and C/O nuclei the differences below ~20-30 MeV/nuc are even more pronounced.

The current data at Voyager runs to July 2013 when the spacecraft is still only ~3.0 AU beyond the location where the heliospheric ACR suddenly disappeared and the new GCR spectra were revealed. So there are many local conditions that could distort the Voyager spectra from true "galactic" spectra. These could include the simple geometrical effects of the proximity of the heliosphere itself; possible extended modulation effects (Scherer, et al., 2011; Strauss, et al., 2013); or the presence of other relatively "local" propagation or source effects that could reduce or increase the intensity at the lowest energies. As Voyager moves deeper into this region and more data, including electrons, heavier primary nuclei such as Mg, Si and Fe and secondary nuclei like 3He, B and N become available, new spectral features and intensity differences, including radial gradients, may become apparent.

Ionization energy loss plays a very important charge dependent role at the lower energies. The observed C/O intensity decrease below ~30 MeV/nuc could be a manifestation of this. We have refrained from examining these effects in detail in this paper for several reasons including the fact that the statistical accuracy that is needed for this study is not yet available.

Another important feature of the Voyager data is that it makes it possible to derive a total modulation function, M=$\beta \ell$n ($j_2/j_1$) for H and He nuclei for the extreme solar minimum conditions in 2009 using the ratios of intensities measured at V1 = $j_2$ and PAMELA = $j_1$. For example, at 200 MeV the simple ratio, R=V1/PAM for H nuclei in Figure 5 is 6.81, for He nuclei in Figure 6 =3.23 (same $\beta$ but different P). These features will be discussed more fully in a subsequent paper.

## 5. Summary and Conclusions

The sudden disappearance of lower energy heliospheric ACR and TSP starting between July 28$^{th}$ and August 25$^{th}$, 2012 when V1 was at 121.7 AU heralded the appearance of spectra and intensities of low energy H, He and C/O nuclei that resembled predictions of these nuclei to be expected in the LIS medium for GCR. This "galactic" component at V1, whose intensity has remained stable to within a few percent now for ~9 months, shows peaks in the differential spectra of these three sets of nuclei in the range 30-60 MeV/nuc.

During this time period both the H and He nuclei spectra can be determined to accuracies ~±5% or better from a few MeV to ~400 MeV/nuc. The accuracy of the C/O data is ~± 10% in each energy interval. The ratio of H nuclei/He nuclei appears to be nearly constant above 30 MeV/nuc with a ratio ~12.5±1.0 in this energy range (see Stone, et al., 2013).

In this paper we are able to reproduce the main features and relative intensities of the observed Voyager spectra using a simple galactic propagation model and with the same "source" spectral index of ~$P^{-2.24}$ for all three sets of nuclei. Using data available at higher energies beyond ~10 GeV/nuc (~20 GV), this exponent appears to remain constant as a function of rigidity from the lowest rigidities to the highest rigidities.

The model used here is a variant of the widely used Leaky Box Model, a simple diffusion model in which the age distribution of the LIS particles can be represented by an exponential. The parameters that are varied to obtain a fit are: 1) The "source" spectrum which is varied from -2.20 to -2.40; 2) The rigidity, $P_0$, at which the diffusion coefficient changes from its higher rigidity dependence of $P^{0.5}$ to an inverse dependence ~$P^{-1.0}$, at the lower rigidities. This results in an increasing value of the diffusion coefficient with decreasing rigidity. This value of $P_0$ is varied from 0.32 to 3.0 GV.

It is found that the measured spectra of H and He nuclei and also C/O nuclei can be reproduced to within 5-10% over the low energy range $\geq$30 MeV measured by Voyager for a source spectral index of -2.24 and a value of $P_0$ between 0.56-0.74 GV. At energies below ~30 MeV/nuc the Voyager data points lie below the calculations, more so for C/O nuclei than for H or He. This agreement between data and calculations extends to higher energies using available high energy data for H and He nuclei up to ~100 GeV/nuc from PAMELA and BESS.

We believe that the differences between the calculations and observations that are seen at lower energies below about 10 MeV for H, about ~20 MeV/nuc for He and 30 MeV nuc for C/O nuclei, could be, in part, related to the uncertainties in the interstellar ionization energy loss of these nuclei. There are no other discernible propagation or source features in the observed Voyager H, He and C/O spectra. We believe that this spectral index =-2.24, constant as a function of rigidity from rigidities ~0.2 GV to at least 200 GeV, is one of the key conclusions that can be obtained from the new Voyager data at 122 AU and beyond.

The ability of such a simple propagation model to reproduce the Voyager observations of the most abundant cosmic ray nuclei including this constancy of the spectra index with such a limited number of propagation parameters in what is obviously a complex galactic environment deserves further consideration, both from the propagation and source distribution/acceleration effects on the spectra of these particles.

**Acknowledgements:**

We greatly appreciate the support of Voyager by JPL. The 2012 data used here comes from Stone, et al., 2013. The earlier Voyager data used here comes from our earlier publications and from internal web-sites provided by Nand Lal and Bryant Heikkila, our co-investigators. The Voyager CRS team is directed by Ed Stone. We appreciate the efforts of all of the team members in compiling the Voyager data.10

# FIGURE CAPTIONS

**Figure 1:** The spectrum of GCR H nuclei measured from 2012.78 to 2013.0 by V1 at an average radial distance of 122.6 AU. All observations here are from Stone, et al., 2013. The black lines in the figure are earlier estimates of the LIS spectrum as per Webber and Higbie, 2009, as described in the text. An earlier spectrum calculated also using a Leaky Box model by Ip and Axford, 1985, is shown as a reference.

**Figure 2:** The same as Figure 1 but for GCR He nuclei observations and earlier calculations.

**Figure 3:** The same as Figure 1 but for GCR C/O nuclei observations and earlier calculations.

**Figure 4:** The interstellar diffusion coefficients originally used in the calculations of Webber and Higbie, 2009. The new limits on this diffusion coefficient which fit the spectra of H and He nuclei measured by V1 (Stone, et al., 2013) that are used in this paper and are shown in red. The sudden change in the rigidity dependence of the diffusion coefficient described by Ptuskin, et al., 2006, with the value of the original diffusion coefficient divided by 2, is shown in blue.

**Figure 5:** The H nuclei spectrum from Voyager shown in Figure 1 along with the new predictions for propagation parameters, $P_0$=0.56, 0.74, 1.00 GV and a ~$\beta^{-1.0}$ dependence of the diffusion coefficient below $P_0$. A source spectrum with exponent = -2.24±0.05 is assumed along with a diffusion coefficient ~$P^{0.5}$ above $P_0$. The data at high energies is from the PAMELA and BESS experiments as described in the text.

**Figure 6:** The Helium nuclei spectrum calculated with the same parameters as in Figure 5. The measured V1 spectrum between 2012.78 and 2013.0 and the PAMELA and BESS measurements at high energies are also shown.

**Figure 7:** The Carbon/Oxygen nuclei spectra calculated with the same parameters as in Figure 5. The measured V1 spectrum between 2012.78 and 2013.0 is also shown. The high energy data is from the surveys of Putze, et al., 2010 and Seo, et al., 2012, and includes the measurements of Englemann, et al., 1990, from ~1-20 GeV/nuc.

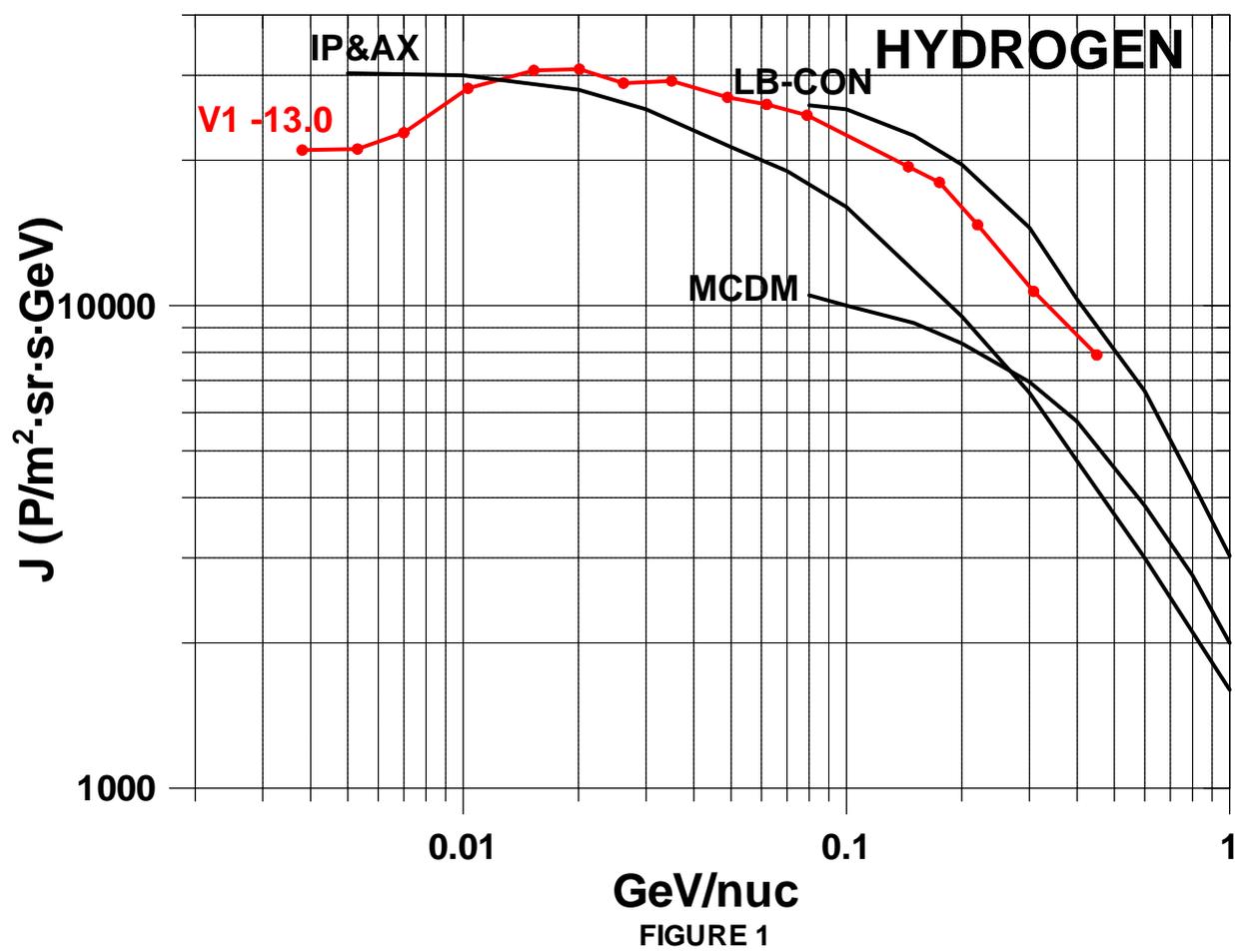

FIGURE 1

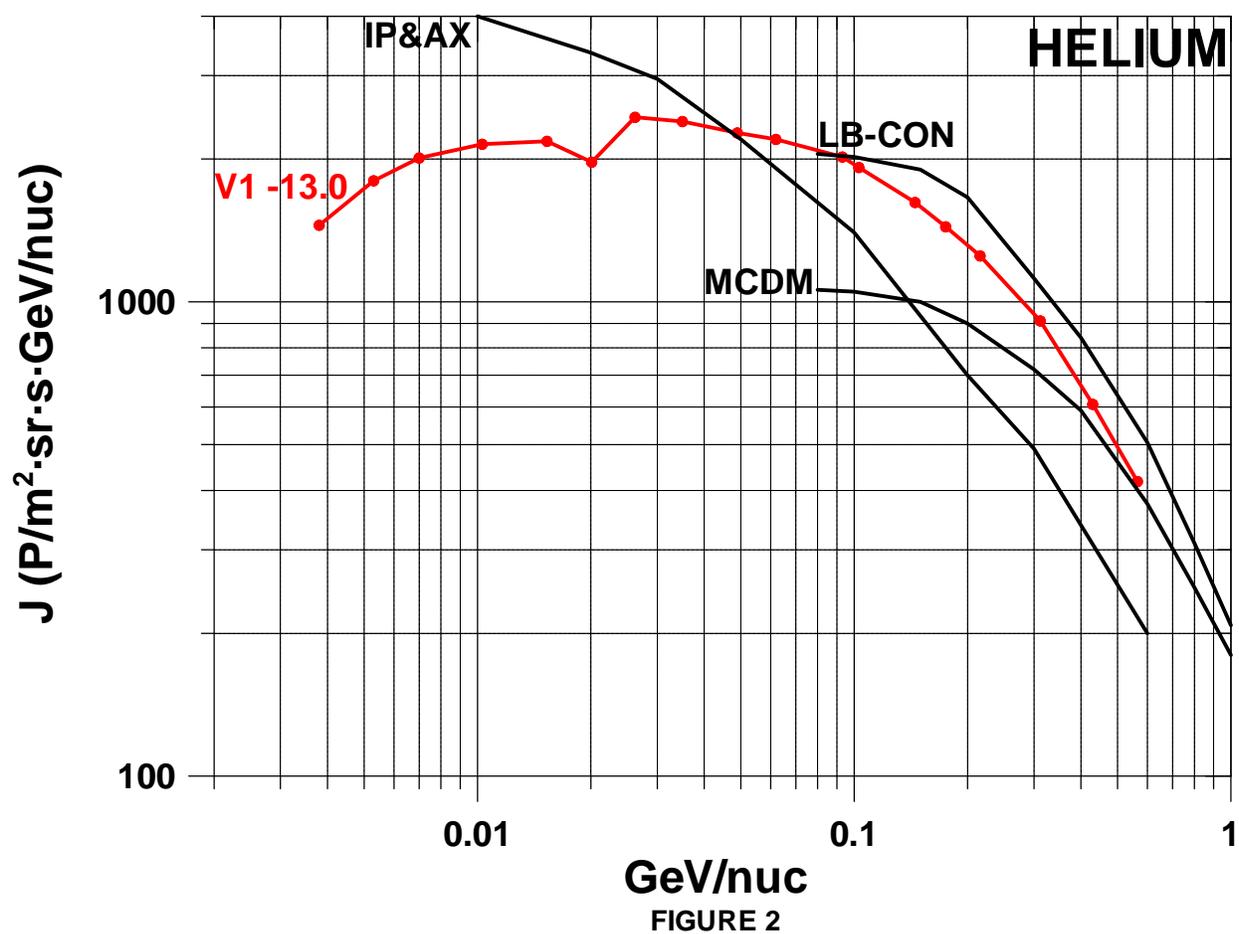

FIGURE 2

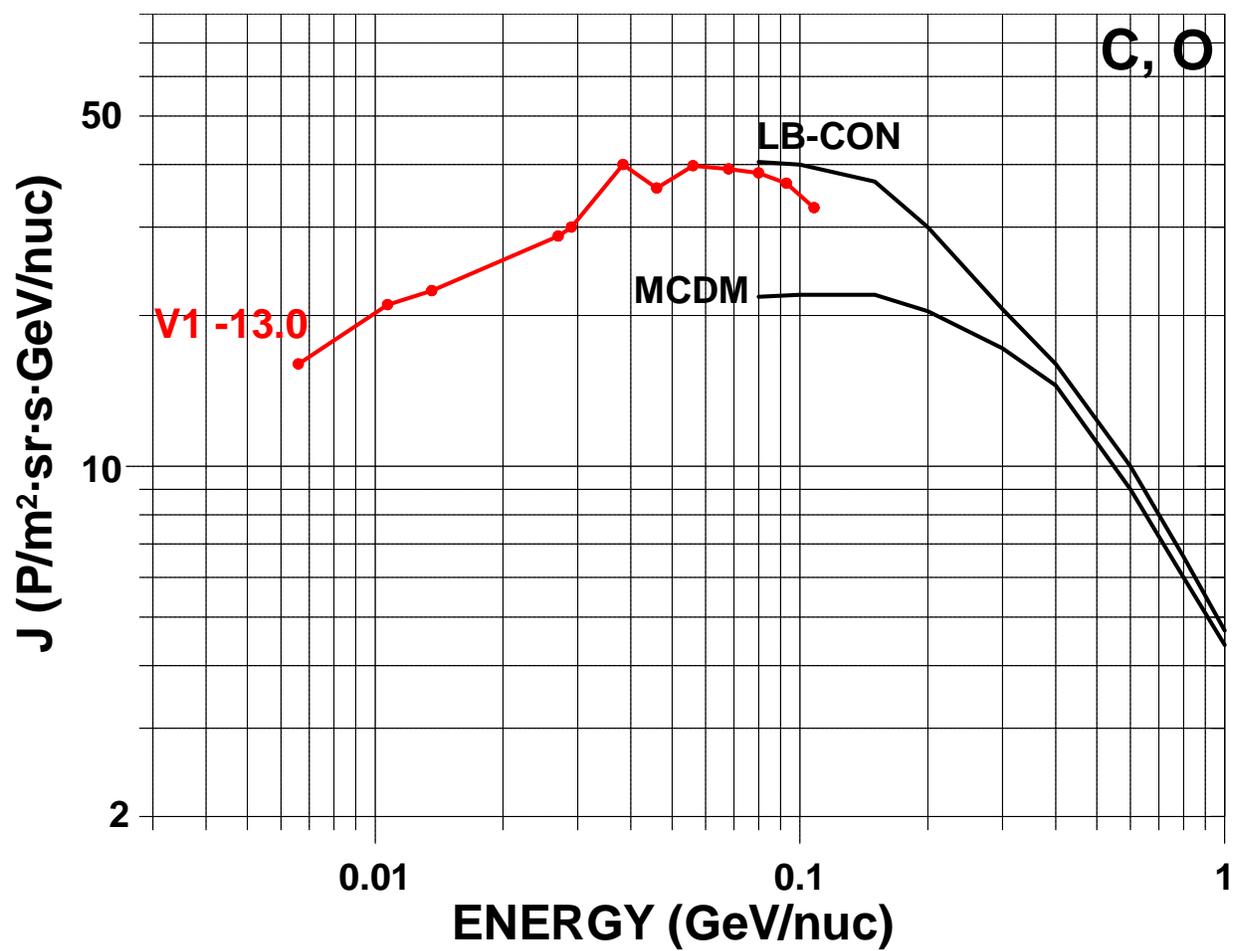

FIGURE 3

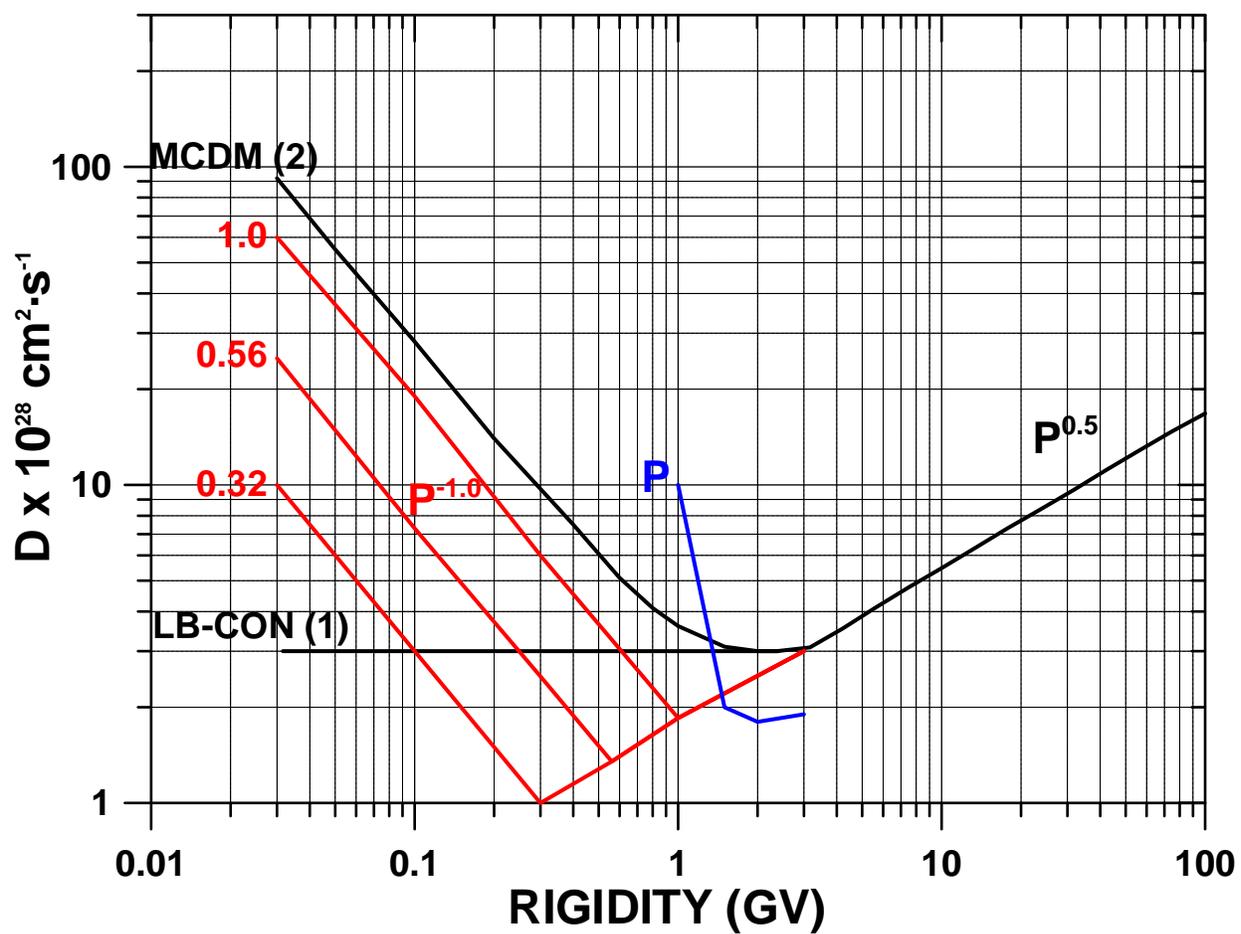

FIGURE 4

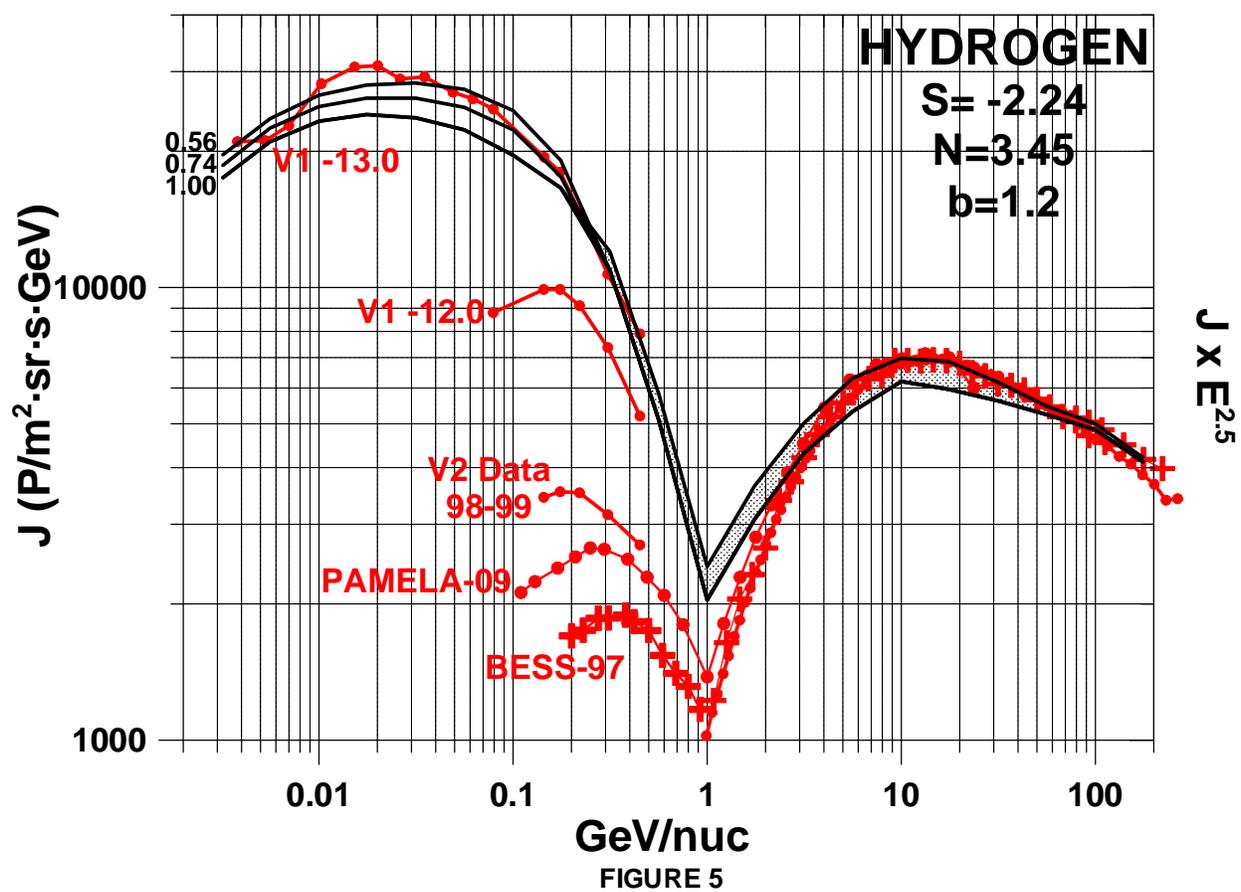

**FIGURE 5**

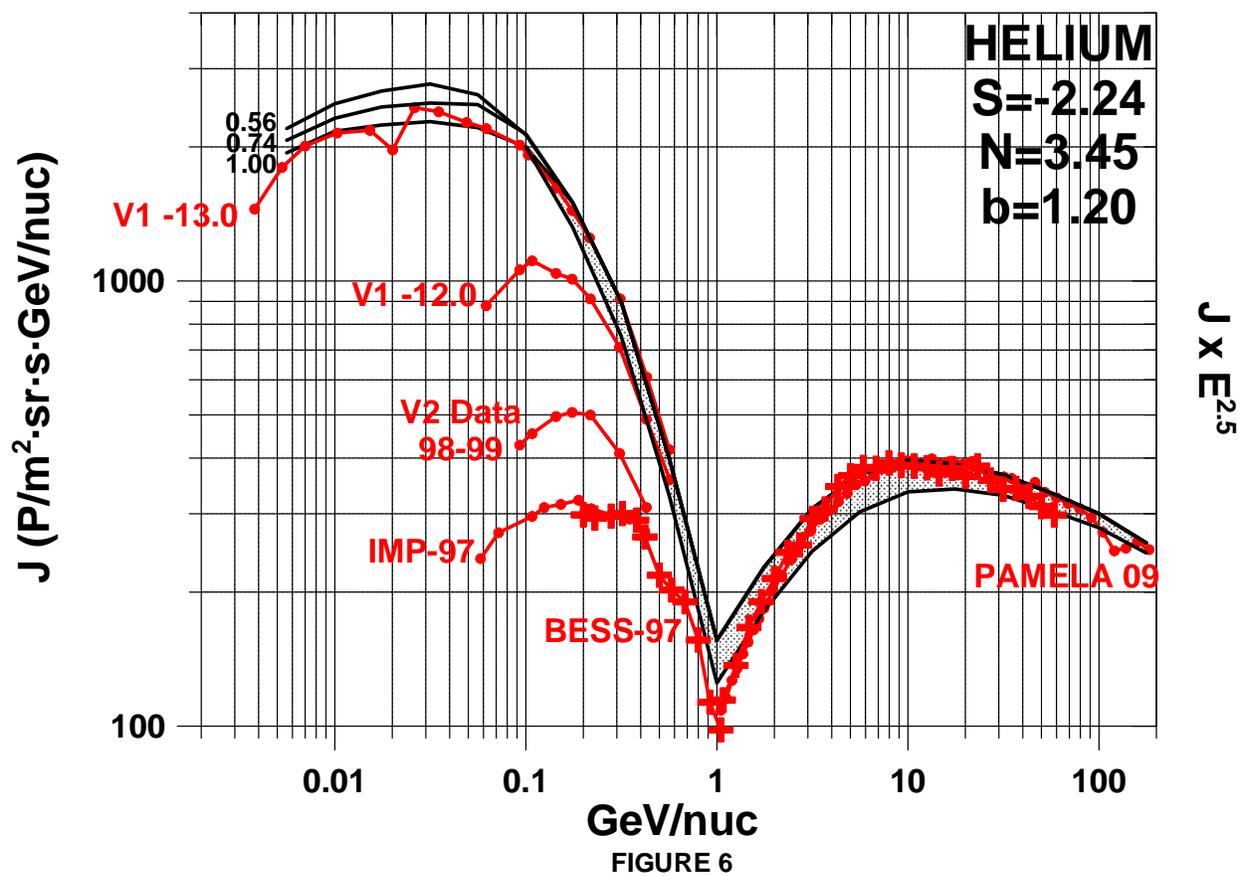

**FIGURE 6**

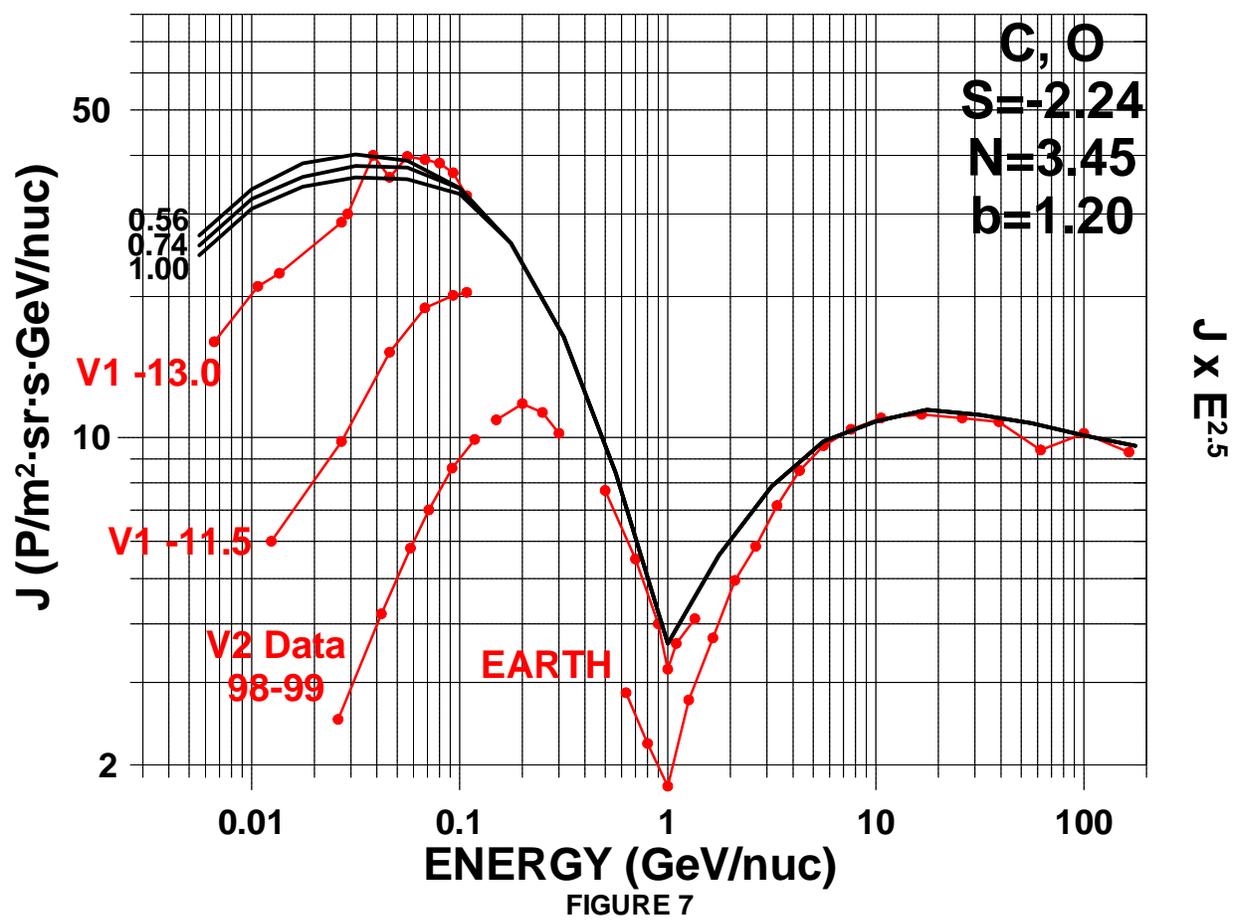

FIGURE 7